\documentclass[twoside,twocolumn,english,aps,prb]{revtex4}
\usepackage[T1]{fontenc}
\usepackage[utf8]{inputenc}
\usepackage{textcomp}
\usepackage{amstext}
\usepackage{graphicx}
\usepackage{setspace}
\usepackage{color}

\makeatletter
\@ifundefined{textcolor}{}
{%
 \definecolor{BLACK}{gray}{0}
 \definecolor{WHITE}{gray}{1}
 \definecolor{RED}{rgb}{1,0,0}
 \definecolor{GREEN}{rgb}{0,1,0}
 \definecolor{BLUE}{rgb}{0,0,1}
 \definecolor{CYAN}{cmyk}{1,0,0,0}
 \definecolor{MAGENTA}{cmyk}{0,1,0,0}
 \definecolor{YELLOW}{cmyk}{0,0,1,0}
}

\makeatother

\usepackage{babel}
\begin{document}

\title{Unusual temperature dependence of the spectral weight near the Fermi
level of NdNiO$_{3}$ thin films}

\author{E. F. Schwier$^{1}$, R. Scherwitzl$^{2}$, Z. Vydrov\`a$^{1}$, M.
Garc\'ia-Fern\'andez$^{1}$, M. Gibert$^{2}$, P. Zubko$^{2}$, M. G.
Garnier$^{1}$, J.-M. Triscone$^{2}$ and P. Aebi$^{1}$}

\affiliation{$^{1}$\textit{\footnotesize D\'epartement de Physique and Fribourg
Center for Nanomaterials, University of Fribourg, CH-1700 Fribourg,
Switzerland}}

\affiliation{$^{2}$\textit{\footnotesize D\'epartement de Physique de la Matière
Condens\'ee, University of Geneva, 24 Quai Ernest-Ansermet, CH-1211
Gen\`eve 4, Switzerland }}
\begin{abstract}
We investigate the behavior of the spectral weight near the
Fermi level of NdNiO$_{3}$ thin films as a function of temperature
across the metal-to-insulator transition (MIT) by means of ultraviolet
photoelectron spectroscopy.  
The spectral weight was found to exhibit thermal hysteresis, similar to that 
of the dc conductivity. A detailed analysis of the temperature dependence reveals 
two distinct regimes
of spectral loss close to the Fermi level. The temperature evolution
of one regime is found to be independent from the MIT.
\end{abstract}
\maketitle

\section{Introduction}

Nickel-based rare-earth perovskite oxides $R$NiO$_{3}$, with $R$
being a trivalent rare earth, are a model system to study temperature-driven
metal-to-insulator transition (MIT). They are among the few oxide
families that exhibit metallic conductivity and, except for LaNiO$_{3}$,
all nickelates undergo an MIT with a critical temperature that depends
on the radius of the rare-earth ion \cite{Torrance:1992p62,Medarde:1997p10}.
The MIT in the nickelates is accompanied by a structural transition
from an orthorhombic (space group Pbnm) \cite{GarciaMunoz:1992p61}
to a monoclinic structure (space group P2$_{1}$/n) \cite{Alonso:1999p78,Staub:2002p34}
with a Ni-O bond length disproportionation. The insulating phase is
thought to exhibit charge order with two inequivalent Ni sites where
the charge from the trivalent ion is split ($d^{n}d^{n}\rightarrow d^{n+\delta}d^{n-\delta}$)
between two neighboring nickel sites \cite{Alonso:1999p69,Alonso:2000p63,Staub:2002p34,Scagnoli:2005p28,Medarde:2009p12}.
For lighter rare earths like Nd and Pr the MIT is accompanied by a
magnetic transition from a paramagnetic to an antiferromagnetic ground
state with unusual spin order. For heavier rare earths the magnetic
transition occurs at lower temperatures than the MIT \cite{GarciaMunoz:1994p72,Scagnoli:2008p30}.

The origin of the metal-insulator transition is currently under debate.
The conduction band of $R$NiO$_{3}$ is constituted by the covalent
bonding of Ni 3d and O 2p orbitals. Initially, nickelates were located
in the Zaanen-Sawatzky-Allen scheme \cite{Zaanen:1985p540} at the
boundary between charge-transfer insulators and low-$\Delta$ metals
\cite{Torrance:1991p1849}. Due to their small charge transfer energy
$\Delta$, it was argued that nickelates may even be considered as
self-doped Mott insulators \cite{Mizokawa:2000p1889}, where electrons
are transferred from the oxygen to the nickel to form a $3d^{8}\underline{L}$
configuration ($\underline{L}$ denotes a hole in the oxygen ligand).
Charge and spin order would then emerge naturally in these compounds
\cite{Mizokawa:2000p1889,Lee:2011p1746}. Alternatively, charge ordering
has been suggested to originate from a strong Hunds coupling, which
would overcome the Coulomb interaction energy $U$ and provide an
alternative route to lift the orbital degeneracy in nickelates \cite{Mazin:2007jx}.
In this picture, nickelates were considered as band insulators rather
than Mott insulators. In recent optical experiments, however, it was
shown that Mott physics is essential to describe the redistribution
of spectral weight \cite{Stewart:2011p1633}. Finally, recent dynamical
mean field theory (DMFT) calculations \cite{Wang:2011p515} suggested
that the Ni occupancy is not compatible with an insulating state caused
by a Mott transition or charge transfer and that new physics beyond
these frameworks might be needed to account for the insulating state
in nickelates.

Ultraviolet photoelectron spectroscopy (UPS) is an ideal tool to probe
the electronic structure and the effects of electronic correlations.
The density of states (DOS) of the valence band is closely related
to the measured photoelectron intensity (spectral weight), and changes
in the latter can be related to the opening of a gap during the MIT
\cite{Mizokawa:1991ef}, variation of the bandwidth \cite{Kang:1997uu}
or the transfer of spectral weight between different parts of the
valence band \cite{Monney:2010vp}. Photoemission measurements on
$R$NiO$_{3}$ have been performed on polycrystalline samples
\cite{Barman:1994p1875,Mizokawa:1995p1844,Medarde:1997p11,Vobornik:1999p65,Okazaki:2003p76}
as well as on thin films \cite{Eguchi:2009p41,Galicka:2009p59,Gray:2011p828,Bilewska:2010p1873}.
Measurements on thin films, however, were only conducted using the
relatively low resolution x-ray photoemission spectroscopy (XPS). The present work 
represents the first high resolution UPS measurements on nickelate thin films.

In recent years the synthesis of nickelate thin films has become increasingly
popular. Due to the absence of sufficiently large single crystals,
epitaxial thin films are the closest system available to study the intrinsic
properties of nickelates. In addition, films can be grown sufficiently
thin to measure the influence of epitaxial strain on the phase transition
\cite{Tiwari:2002dv,DeNatale:1995p47,Catalan:2000p9,Novojilov:ct,Tiwari:2002p52},
opening another degree of freedom in the exploration of the phase
diagram of nickelates.

In this paper we present an analysis of the temperature dependence
of the spectral weight near the Fermi level of NdNiO$_{3}$ thin films
measured with UPS. We compare measurements at different temperatures
from well above to well below the phase transition. We study changes
in the valence band as well as the loss of spectral weight near the
Fermi edge during the transition. The absence of a complete gap opening
in the system is investigated by an analysis of the temperature dependent
evolution of the spectral weight in the vicinity of the Fermi level.
Finally, the loss of spectral weight is separated into two regimes:
at and close to the Fermi energy, and reasons for their different-temperature
evolutions are discussed.

\section{Experiment}

Epitaxial NdNiO$_{3}$ thin films were grown on (001)- oriented Nb:SrTiO$_{3}$
and SrTiO$_{3}$ substrates by radio-frequency off-axis magnetron
sputtering in 0.180 mbar of an oxygen/argon mixture of ratio 1:3 at
a substrate temperature of 490\textdegree{}C. More details on the
growth conditions and structural characterization of NdNiO$_{3}$
films can be found in Ref. \cite{Scherwitzl:2010p4}. In this study,
the film thickness was approximately 100 nm. 
DC transport measurements were carried out in a four-point configuration 
between room temperature and 4.2 K in a helium dipping station with a 
heating/cooling rate of approximately 1 K/min.
Four platinum electrodes were deposited in each corner
of the sample and the resistance R (or conductance $\sigma=1/R$)
was measured in constant current mode. The transition temperature
$T_{MIT}$ is defined as the temperature where $d(\ln R)/dT$ is maximum. 

The ex-situ transferred films were cleaned from adsorbates by annealing
in an oxygen atmosphere ($P_{O_{2}}\approx200$ mbar) at $T\approx750$
K for more than two hours. The surface cleanliness was tested using
x-ray photoelectron spectroscopy of the C 1s core levels, whereas
the quality of the atomic order at the surface could be verified by
means of low electron energy diffraction. 

The UPS spectra were recorded using an upgraded SCIENTA SES 200 analyzer.
S-polarized monochromatic photons from the He I$_{\alpha}$ line ($h\nu=21.2$
eV) of a helium discharge lamp were used to excite the photoelectrons.
The base pressure in the chamber did not exceed 4$\times$10$^{-11}$
mbar, while the He partial pressure during the measurement was below
5$\times$10$^{-10}$ mbar. All spectra were recorded at normal emission
with angular integration and with an energy resolution better than
8 meV. The position of the Fermi energy ($E_{F}$) was determined
by fitting the spectral weight of sputtered polycrystalline Nb at
a temperature of 17 K. The constant photon flux was verified manually
during the measurements and the spectra are not normalized. The results
were reproduced on different samples during multiple experiments.

Sample cooling and heating during the photoemission experiments was
achieved by thermal contact with LHe and resistive heating of the
manipulator at a rate of less than 2 K/min for the high resolution
heating ramp and less than 4 K/min during the complete cooling and
heating cycle. The samples remained at the lowest temperature for
at least one hour before the heating commenced.

\section{Results \& Discussion}

\subsection*{The valence band above and below T$_{MIT}$}

\begin{figure}[h]
\begin{centering}
\includegraphics{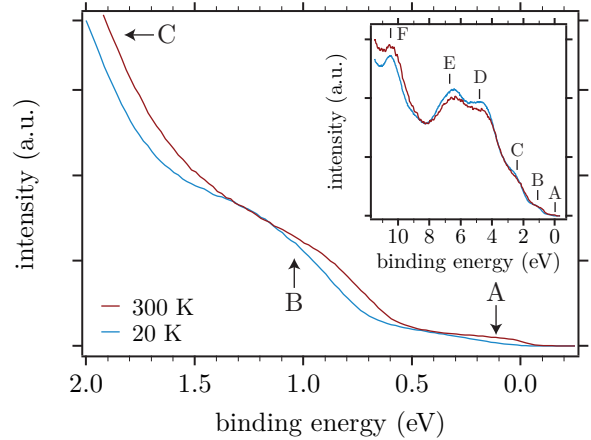}
\par
\end{centering}
\caption{Region of the valence band near $E_{F}$ above (red) and below (blue)
the MIT as measured with He I$_{\alpha}$. At $E_{F}$ a loss of spectral
weight is present (A), while at higher binding energies features in
the spectral weight are shifted by approximately 0.2 eV (B \& C).
(Inset) The full valence band of NdNiO$_{3}$ at the same temperatures.
Black bars mark the peak positions as determined from a second derivative
of the low temperature spectrum.\label{fig_01_VB}}
\end{figure}

In Fig. \ref{fig_01_VB} the valence band of NdNiO$_{3}$ near $E_{F}$
is plotted for two temperatures. The spectrum at 20 K (blue) corresponds
to the antiferromagnetic insulating phase while the spectrum at room
temperature (red) corresponds to the film being in the paramagnetic
metallic phase. The inset shows the full valence band plotted for
the same two temperatures. It is composed of several features (A-F). 

The $A$ region has been associated with the $e_{g}^{*}$ band ($E_{B}<400\,\textrm{{meV}}$).
It is situated in close proximity to $E_{F}$ and its character is
dominated by the strong hybridization of oxygen and nickel states
\cite{Sarma:1994p2228,Barman:1994p1875,Mizokawa:1995p1844}. Upon
cooling the $A$ region exhibits a strong loss of spectral weight
from the metallic to the insulating phase. Such a behavior is consistent
with previous photoemission measurements and has been linked to changes
in the initial and final state configuration as well as to an opening
of a charge-transfer gap \cite{Medarde:1997p11,Vobornik:1999p65,Okazaki:2003p76,Eguchi:2009p41,Galicka:2009p59}.

The $B$ peak ($E_{B}\approx1$ eV) can be associated with the $t_{2g}^{*}$
band of NdNiO$_{3}$. \emph{Ab initio} calculations \cite{Barman:1994p1875,Sarma:1994p2228}
propose a similar strength of orbital hybridization as in the $A$
region. At lower temperatures, this feature is shifted by approximately
0.2 eV to higher binding energies. This is compatible with similar
changes reported by Eguchi et \emph{al.} who performed photoemission
with a bulk-sensitive excitation energy of $h\nu=7927$ eV \cite{Eguchi:2009p41}.
The nature of this shift might be related to the transfer of spectral
weight found in the manganite Pr$_{1/2}$Sr$_{1/2}$MnO$_{3}$ during
the MIT \cite{Chainani:1997ew}, where a comparable shift of 0.2 eV
is present.

The $C$ peak ($E_{B}\approx2.4$ eV) is expected to be composed of
O 2p and, to a smaller part, of Ni 3d states \cite{Barman:1994p1875,Sarma:1994p2228}.
Like the $B$ peak, it also shifts by approximately 0.2 eV to higher
binding energies. The comparable shifts of both the $B$ and the $C$
peaks suggest a similar driving force coupling to the binding energy
of both features. A precise analysis of the shift of both bands as
a function of temperature should provide additional information regarding
this hypothesis.

The $D$ peak ($E_{B}\approx4.8$ eV) and $E$ peak ($E_{B}\approx6.7$
eV) are considered to be dominated by O 2p states, where, in the case
of the $D$ peak, localized Nd 4f states also contribute to the spectral
weight \cite{Sarma:1980p2176,Barman:1994p1875}. At low temperatures
both peaks slightly increase in intensity. It seems unlikely that
this increase is related to additional adsorption of oxygen at the
surface, as x-ray photoemission measurements of the O 1s core level
do not show any increased emission at low temperatures. It should
be noted that the $E$ peak does exhibit a higher intensity compared
to what is known from literature. A comparison of different publications
shows that this feature is more prominent in thin films \cite{Galicka:2009p59,Eguchi:2009p41}
than in polycrystalline samples \cite{Barman:1994p1875,Medarde:1997p11,Vobornik:1999p65}.
This suggests, that the highly coordinated single-crystalline lattice
of the thin film samples is a possible reason for this difference.

The $F$ peak ($E_{B}\approx10.5$ eV) can be identified, by comparison
with PrNiO$_{3}$, to be the result of a photoemission satellite with
$d^{6}$ and $d^{8}\underline{L}^{2}$ character \cite{Mizokawa:1995p1844}.
The suggestion by Galicka et \emph{al.} \cite{Galicka:2009p59} that
a peak at this binding energy might be related to residual carbon
at the surface of the sample can be ruled out in the present case
due to the absence of C 1s in the XPS measurements. It should be noted
that the peak is broadened towards higher binding energies in the
insulating phase. Its satellite nature connects this broadening to
a change in the electronic configuration of the valence electrons
at $E_{F}$.

\subsection*{Energy distribution curves at different temperatures }

In this paper the focus is set on the strong reduction of spectral
weight in the $e_{g}^{*}$ band ($A$ region) crossing $E_{F}$. To
investigate the intensity loss and its temperature dependence in this
region, several high resolution spectra were recorded while the sample
was heated from 10 K to ambient temperature.

\begin{figure}[h]
\begin{centering}
\includegraphics{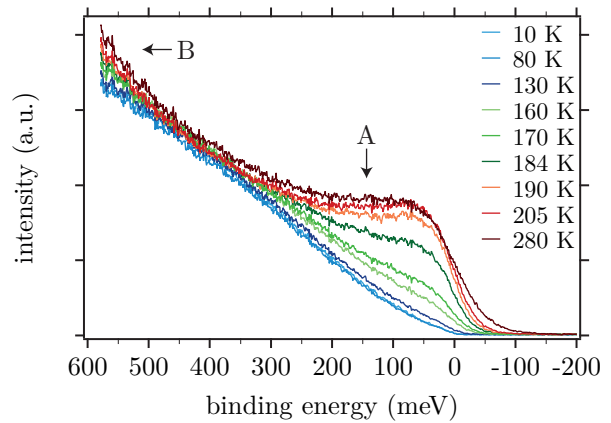}
\par\end{centering}

\caption{Angle-integrated energy distribution curves of the spectral weight
close to $E_{F}$. The MIT is accompanied by an almost complete suppression
of spectral weight at $E_{F}$ as well as a loss of spectral weight
at higher binding energies. Increased loss at 600 meV is caused by
the shift of bands around 1 eV (see Fig.\ref{fig_01_VB}). \label{fig_02_eF}}
\end{figure}

In Fig. \ref{fig_02_eF} the raw energy distribution curves (EDCs)
recorded close to $E_{F}$ at different temperatures have been plotted.
The spectra were obtained during a heating cycle to ensure the fully
insulating state of the sample at low temperatures \cite{Kumar:2009gj}.
A comparison between the highest ($T=1.60T_{MIT}$) and lowest ($T=0.06T_{MIT}$)
temperature spectra shows a strong decrease of the spectral weight
($A$ region) between binding energies of approximately 400 meV and
$E_{F}$. Also visible is a region above 450 meV which exhibits a
temperature dependent loss. The latter is, in fact, the tail of the
$B$ peak shifting to higher binding energies during the MIT (see
Fig. \ref{fig_01_VB}). In order to highlight the evolution of spectral
weight in the $A$ region, the EDCs will be separated into two different
temperature regimes: a low temperature regime where the sample is
in a nominally insulating state, and a temperature regime around the
MIT where coexistence of metallic and insulating domains is expected.

The spectra at temperatures up to 80 K do not exhibit any temperature
dependence and between 80 K and 130 K only a very weak intensity increase
is present. All spectra show a small but finite contribution of spectral
weight at and above $E_{F}$ without a resolvable Fermi edge. Also,
an almost linear slope is present in the energy-dependent intensity
across the whole $A$ region. 

The fact that no Fermi edge can be resolved in the low temperature
spectra supports the notion of a fully insulating sample without metallic
domains. However, there is no complete gap at $E_{F}$ either, as
demonstrated by the residual intensity present even at lowest temperatures.
The presence of a gap is expected in the insulating state of a normal MIT. Its
absence has to be interpreted as a sign of the unusual nature of the
transition.

It has been proposed that the abscence of a complete gap in the spectra
may be the result of a highly assymetric gap broadended by experimental
resolution \cite{Vobornik:1999}. The valence band onset in such a
system would lie very close to but below $E_{F}.$ The measurement
process then leads to a convolution between the spectral function
including the gap and the instrumental broadening. This, in turn,
would result in a residual intensity at $E_{F}$ in the low temperature
spectra. 

In Fig. \ref{fig_02_eF} the onset of the resolution-broadened valence
band can be estimated to lie between ($-5\pm3$) meV (10 - 80 K) and
($-25\pm3$) meV (130 K) in the insulating regime. The onset at the lowest temperature
 is compatible with our experimental resolution (8
meV). However, the onset at 130 K ($-25$ meV) cannot be explained
by resolution broadening, meaning that the system cannot exhibit a
complete gap in the insulating phase. This means that the loss of
spectral weight has to be caused by a different process. There are
several possible explanations for the observed evolution of the spectral
weight.

Firstly, the MIT has to be understood as a transition from a metal
to a semiconductor with a very low activation energy \cite{Granados:1993p49,Catalan:2000p7}.
In this scenario there is a completely gapped Fermi surface but localized
donor or acceptor states are present within the gap. These states,
given a high enough density, can contribute to the total photoemission
signal with a finite intensity in the photoemission spectra at $E_{F}$
\cite{Higuchi:1998wv,Aiura:2002vh}. Also, a polaronic in-gap state
has been proposed to be present in La$_{1/3}$Sr$_{2/3}$FeO$_{3}$
\cite{Wadati:2006up} after the system undergoes a transition from
a metallic to an insulating state with charge and spin order similar
to NdNiO$_{3}$.

Secondly, in a manner similar to d-wave superconductors \cite{Damascelli:2003kq}
or charge density wave compounds \cite{Brouet:2004fl}, only parts
of the Fermi surface might actually be gapped, leaving bands crossing
$E_{F}$ to contribute to the spectral weight. The idea of a partially
gapped Fermi surface is given support by the fact that optics experiments
\cite{Katsufuji:1995p43,Stewart:2011p1633}, indicate a gap
opening in NdNiO$_{3}$ while only probing the band structure at the
center of the Brillouin zone. In the presence of a Mott type
transition this could lead to a zero-size gap \cite{Fazekas:1980un},
where residual density of states remains at temperatures lower than
$T_{MIT}$ due to an overlap of the upper and lower Hubbard bands
at $E_{F}$ \cite{Thouless:1976hx}. Dardel et \emph{al.} \cite{Dardel:1992ug}
were able to fit residual spectral weight below $T_{MIT}$ in the
Mott insulator TaS$_{2}$ using this framework. Similar fits,
however, do not converge for the present data.

Another fascinating possibility was presented through calculations
of Lee et \emph{al.} \cite{Lee:2011p1746} by an application of DMFT.
They argued, that a bond-centered spin-density-wave state could be
present in the low temperature phase of the nickelates. This leads
to Dirac-cone-like bands crossing $E_{F}$ with a linear density of
states that would resemble the linear spectral weight present in our
data. They predict the system to be semi-metallic in the presence
of charge order. A gap is only formed if the orthorhombic distortion
of the crystal becomes sufficiently high, which might be the case
for heavier rare earths. So far, the hypothesis that a gap might emerge
in the electronic structure under sufficiently large orthorhombic
distortion could not be confirmed by photoemission measurements on
EuNiO$_{3}$ and SmNiO$_{3}$ \cite{Bilewska:2010p1873,Vobornik:1999}. 

\begin{figure}[th]
\begin{centering}
\includegraphics{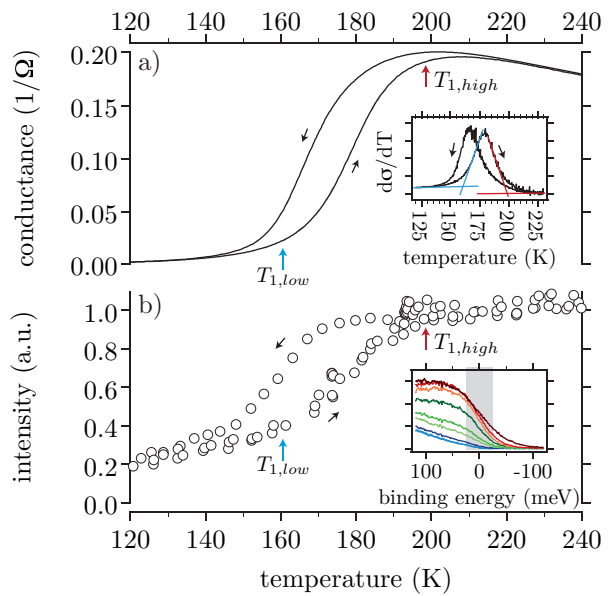}
\par\end{centering}

\caption{a) Conductance of NdNiO$_{3}$ as a function of temperature. The blue
and red arrows mark the two critical temperatures of the hysteresis
on heating, which are defined by the intersections of the blue and
red lines in the inset. (Inset) Temperature derivative of the conductance.
b) The integrated spectral weight at $E_{F}$ as a function of temperature.
The plot was normalized to the constant intensity above 250 K. (Inset)
Energy distribution curves from Fig. \ref{fig_02_eF} with the $\pm$
25 meV interval around $E_{F}$ used for integration. \label{fig_03_hysteresis}}
\end{figure}

At higher temperatures ($T>160$ K), a Fermi edge becomes clearly
visible in the spectra (Fig. \ref{fig_02_eF}) and can be fitted using
a polynomial density of states. These fits result in a reasonable
effective temperature as well as the correct position of $E_{F}$.
The emergence of emission from metallic domains is to be expected
at temperatures above the critical temperature of the phase transition $T_{1,low}$
(see Fig. \ref{fig_03_hysteresis} a). In addition to the emergence
of a Fermi edge, the region between the two critical temperatures
(see Fig. \ref{fig_03_hysteresis} a) of the MIT ($T\approx160-200$
K) is marked by a sudden increase of the spectral weight that contributes
approximately 2/3 of the total loss at $E_{F}$. These results are
generally in good agreement with previous photoemission experiments
on polycrystalline $R$NiO$_{3}$ crystals. \cite{Medarde:1997p11,Vobornik:1999p65,Okazaki:2003p76,Eguchi:2009p41}. 

So far, the analysis of the EDCs strongly indicates that no complete
gap is present in the electronic structure. It is, however, not clear
if the loss of spectral weight in region $A$, as well as the residual
intensity at $E_{F}$ in the low temperature spectra, are the result
of a pseudo gap, intensity from a tail of lower-lying O 2p states,
donor levels inside the gap, a novel electronic state at low temperatures
or a combination of the above. An answer to this question could be
obtained by future angle resolved measurements which are able to resolve
the dispersion of the electronic bands above and below the MIT.

\subsection*{Comparison of spectral weight and conductivity}

Even in the absence of momentum resolved spectra it is still possible
to shed further light on the behavior of the electronic states at
$E_{F}$ by comparing the evolution of the effective number of charge
carriers with the actual conductivity of the sample. The presence
of any kind of gap in the electronic structure is expected to result
in a correlation between the transport data and the integrated spectral
weight at the Fermi level \cite{Dardel:1992ea}.

In Fig. \ref{fig_03_hysteresis}, the thin film conductance (Fig.
\ref{fig_03_hysteresis} a) is compared to the integrated spectral
weight at $E_{F}$ (Fig. \ref{fig_03_hysteresis} b) as a function
of temperature in a window around $T_{MIT}$. The top inset shows
the temperature derivative of the conductance used to define the
critical temperatures of the transition $T_{1,low}$ \& $T_{1,high}$.
The bottom inset illustrates the energy window used to obtain the
integrated spectral weight. 

The conductance (Fig. \ref{fig_03_hysteresis} a) exhibits a narrow
hysteresis ($\Delta T\approx13$ K) with the MIT temperatures of $T_{MIT}=T_{MIT,\uparrow}=179$
K on heating and $T_{MIT,\downarrow}=166$ K on cooling (inset in
Fig. \ref{fig_03_hysteresis} a). To quantify the extent of the MIT,
we define the critical temperatures $T_{1,low}$ and $T_{1,high}$
for the heating branch. These are estimated as shown in the top inset
of Fig. \ref{fig_03_hysteresis} a) to be $T_{1,low}\approx160$ K
and $T_{1,high}\approx200$ K from the intersection of the blue and
red lines respectively. These temperatures mark sharp changes in the
conductivity data and thus roughly quantify the boundaries for the
phase coexistence region or percolation thresholds for the metallic
and insulating states.

The integrated photoemission intensity in Fig. \ref{fig_03_hysteresis}
b) is also hysteretic, with critical temperatures that
are comparable \cite{footnote:temperature} to the values determined
from transport. 
By analogy with the thermal hysteresis in conductivity 
\cite{Catalan:2000p7}, the observed temperature hysteresis 
in the spectral weight implies the coexistence of metallic and semiconducting 
domains during the transition, both contributing individually to the photoemission signal. The determination
of $T_{MIT}$ by means of an analysis of the integrated spectral weight
has been successfully performed on polycrystalline PrNiO$_{3}$ \cite{Medarde:1992p13}.
In this work, however, it was possible for the first time to resolve the hysteresis in these systems.

It should be noted that hysteretic behavior has been reported
for several other system parameters associated with the MIT. Among
these are the temperature dependence of the magnetic moment at the
nickel site \cite{Medarde:1997p10,Caytuero:2006p50,Scagnoli:2006p31},
the unit cell volume \cite{Ruello:2005p45}, as well as transport
properties like the Seebeck coefficient and the resistivity \cite{Granados:1992p1850}.
The hysteresis in the photoemission data is a demonstration that the
MIT is indeed responsible for the loss of spectral weight at $E_{F}$.

\subsection*{Temperature-dependent loss of spectral weight}

Since the loss of spectral weight at $E_{F}$ can be connected to
the MIT, we will now perform a closer analysis of its temperature
and binding energy dependence in the $A$ region.

\begin{figure}[h]
\begin{centering}
\includegraphics{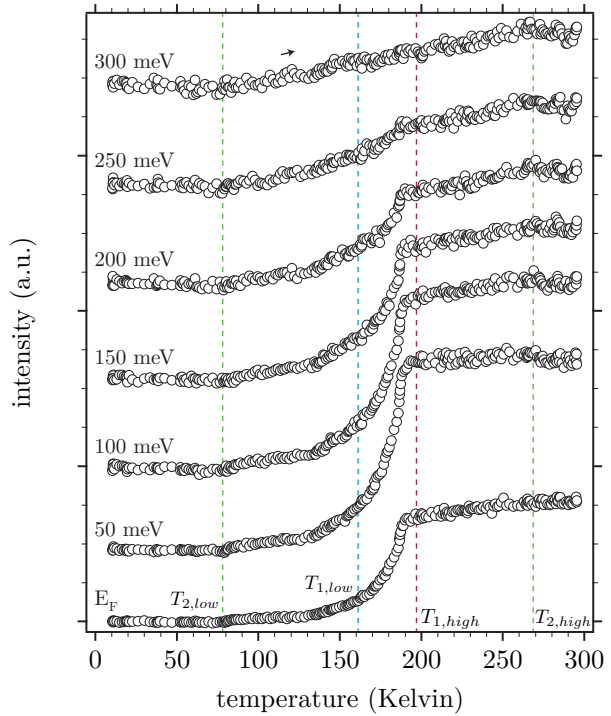}
\par\end{centering}

\caption{Spectral weight integrated over a $\pm$ 25 meV window around various
binding energies close to $E_{F}$ as a function of temperature while
heating the sample. The blue and red dashed lines denote the critical temperatures
of the MIT on heating ($T_{1,low}$ \& $T_{1,high}$) as obtained
from the transport data. The green dashed lines ($T_{2,low}$ \& $T_{2,high}$)
act as guides to the eye for the changes in slope in the TDCs around
80 and 270 K, most prominent at higher binding energies.\label{fig_06_ADCs}}
\end{figure}

In Fig. \ref{fig_06_ADCs} the integrated intensity at different binding
energies within the $A$ region has been plotted as temperature distribution
curves (TDCs) obtained during a heating ramp. The intensity was integrated
within windows of $\pm$ 25 meV around the annotated binding energies.
A variation of this integration window between $\pm$ 3 meV and $\pm$
100 meV did not lead to a qualitative change of the plot. The curves
have been offset against each other for clarity as the room temperature
intensities are similar over a broad range of binding energies.

In the vicinity of $E_{F}$ the strongest feature is the sudden increase
of intensity around $T_{MIT}$ between the 
critical temperatures of the MIT $T_{1,low}$ and $T_{1,high}$ (blue and red dashed
lines). This feature completely disappears at higher binding energies
where only a linear increase of the spectral weight is present over
a broad temperature range (see 300 meV curve). 

The sudden increase of intensity around the MIT, as well as the absence
of a similarly abrupt intensity change in the high binding energy
region of the spectral loss, strongly suggest that the overall temperature
evolution within the $A$ region is actually governed by two regimes
($A_{1}$ \& $A_{2}$) with distinct temperature dependences.

The fingerprint of the first regime $A_{1}$ is the sharp rise of
intensity during the MIT between 160 and 200 K ($T_{1,low}$ \& $T_{1,high}$
in Fig. \ref{fig_06_ADCs}). It is most prominent around $E_{F}$
and its contribution to the total spectral loss vanishes almost completely
around binding energies of 250 meV. The correlation between the intensity
loss of $A_{1}$ close to $E_{F}$ and the MIT was already demonstrated
during the comparison of the integrated spectral weight at $E_{F}$
with the conductivity of the sample (see Fig. \ref{fig_03_hysteresis}).
Further support is given by the comparison of the temperature dependence
of $A_{1}$ with the effective number of charge carries determined
from the integration of the optical conductivity by Katsufuji et \emph{al.}
\cite{Katsufuji:1995p43}. They observed a sudden increase in effective
carrier density around 150 K and a saturation at 200 K, which is consistent
with the temperature dependence of $A_{1}$ observed in this work,
despite the lower temperature resolution of the optical conductivity
measurements. It is therefore evident that $A_{1}$ is directly connected
to the physics of the MIT. 

The second regime $A_{2}$ is governed by the linear increase of intensity
over a broad temperature range. The intersections of linear fits are
at $T_{2,low}\approx80$ K and $T_{2,high}\approx270$ K for 
 characteristic temperatures of the linear increase at a binding energy
of 300 meV (green dashed lines). These temperatures are, in fact,
found to mark a change of the intensity evolution in all TDCs up to
$E_{F}$. The similarity of the characteristic temperatures in TDCs at all binding energies gives
support to the idea that the $A_{2}$ loss is not only confined to higher
binding energies ($E_B\approx300$ meV), but that $A_{2}$ also governs the temperature
evolution at $E_{F}$ within its respective temperature range together with $A_{1}$\cite{footnote:intensity}.

A comparison of the characteristic temperatures of $A_{2}$ ($T_{2,low}$ \&
$T_{2,high}$) with  those of the MIT ($T_{1,low}$,
$T_{MIT}$ \& $T_{1,high}$) shows that changes in the TDC that are
attributed to $A_{2}$ are located far away from the MIT and therefore cannot
be linked to the phase transition. A connection between $A_2$ and $B$ can also 
be ruled out, as the temperature dependence of the $B$ region is similar to that of $A_1$\cite{footnote:BA2}. 
Another yet unknown process has
to govern the temperature evolution of $A_{2}$. In this regard it
is interesting to note that both temperatures ($T_{2,low}$ \& $T_{2,high}$)
are in close proximity to the temperatures reported to be related to changes
in the magnetic properties of the system. At a temperature comparable to $T_{2,low}$ for instance
, Mallik et \emph{al.} \cite{Mallik:1998p1876} found a
change in the sign of the magnetoresistance of NdNiO$_{3}$. At temperatures higher
than $T_{MIT}$ in the itinerant phase, the evolution of the magnetic
moment at the Ni site has been extrapolated to yield a virtual Néel
temperature at around 250 K (neutron diffraction) \cite{Vobornik:1999,GarciaMunoz:1994p72}
or 230 K (Mössbauer effect) \cite{Caytuero:2006p50}. The latter study
also proposed magnons to be present at least in the low temperature
regime of the system. The several hundred meV binding energy range
of the $A_{2}$ regime is also compatible with the range of magnon
dispersion \cite{Delannoy:2009p1877}. Also, the presence of a virtual
Néel temperature has been suggested \cite{Catalan:2008p8} to be reflected
in a change of slope in the susceptibility measured by Zhou et \emph{al.}
\cite{Zhou:2000p1841} hinting at a magnetic transition in the itinerant
phase. 

Although no direct proof of a correlation between magnetic 
properties and electronic structure in the nickelates can be obtained from our data, 
the possibility that the observed loss $A_{2}$ is of magnetic origin certainly merits further study. 
The correlation between spectral weight and magnetic fluctuations has already been 
discussed by Kampf et \emph{al.} for the metallic pseudogap phase of the cuprates \cite{Kampf:1990vu}.
Future measurements should therefore focus on changes in the electronic structure around 
these temperatures. Employing techniques other than photoemission, which are sensitive 
to the magnetic properties of the system, would provide usefull information.

\section{Conclusion}

By measuring the temperature evolution of the valence band of NdNiO$_{3}$
thin films, we were able to highlight the subtle changes in the electronic
structure of NdNiO$_{3}$ thin films as a function of temperature.
We could further demonstrate the presence of thermal hysteresis in
the spectral weight at the Fermi level and relate its evolution to
the conductivity of the thin film. The large temperature range investigated
together with the high temperature and energy resolution allowed us
to distinguish two different loss regimes within the spectral weight
close to $E_{F}$. 

The first regime could be directly related to changes induced by the
MIT. Several possible mechanisms responsible for the MIT and the existence
of residual intensity at $E_{F}$ in the insulating phase were discussed.
By analyzing the temperature evolution of the spectral weight at various
binding energies, a second regime could be identified, which was shown
to be completely disconnected from the MIT, with  characteristic temperatures
for the loss of spectral weight far above and below $T_{MIT}$. After
comparison with measurements sensitive to magnetic structure, it was
suggested that changes in the magnetic state of the system could be
responsible for this interesting temperature behavior.

$\,$
\begin{acknowledgments}
The authors would like to thank A. J. Millis, D. Baeriswyl, V. K.
Malik, C. Battaglia and C. Didiot for helpful discussions as well as the mechanics
and electronics workshops in Fribourg and Geneva for their support.

This project was supported by the Fonds National Suisse pour la Recherche
Scientifique through Division II and the Swiss National Center of
Competence in Research MaNEP.
\end{acknowledgments}
\bibliographystyle{apsrev}

\begin{thebibliography}{63}
\bibitem[{foo()}]{footnote:temperature}
\bibinfo{note}{Note the measured temperatures used for the photoemission data
  are offset from the actual sample temperatures due to the difference in heat
  conductance between LHe-thermocouple and LHe-sample, respectively}.
\bibitem[{foo()}]{footnote:intensity}
\bibinfo{note}{A change of intensity due to de- and adsorbtion of residual gas layers 
during the temperature ramp can be ruled out as temperature dependent desorbtion 
processes are typically confined to much smaller time and temperature scales due to the activated 
nature of the desorption.}
\end{thebibliography}

\end{document}